\input{aipcheck}
\documentclass[final]{aipproc}
\layoutstyle{6x9}

\begin{document}

\title{Coherent Patterns in Nuclei and in Financial Markets}


\author{S.~Dro\.zd\.z}{
 address={Institute of Nuclear Physics, Polish Academy of Science, PL--31-342 Krak\'ow, Poland},
  altaddress={Faculty of Mathematics and Natural Sciences, University of Rzesz\'ow, PL--35-310 Rzesz\'ow, Poland}
}

\author{J.~Kwapie\'n}{
 address={Institute of Nuclear Physics, Polish Academy of Science, PL--31-342 Krak\'ow, Poland}
}

\author{J.~Speth}{
 address={Institut f\"ur Kernphysik, Forschungszentrum J\"ulich, D-52425 J\"ulich, Germany}
}

\begin{abstract}
In the area of traditional physics the atomic nucleus belongs to the most complex systems. It involves essentially all elements that characterize complexity including the most distinctive one whose essence is a permanent coexistence of coherent patterns and of randomness. From a more interdisciplinary perspective, these are the financial markets that represent an extreme complexity. Here, based on the matrix formalism, we set some parallels between several characteristics of complexity in the above two systems. We, in particular, refer to the concept - historically originating from nuclear physics considerations - of the random matrix theory and demonstrate its utility in quantifying characteristics of the coexistence of chaos and collectivity also for the financial markets. In this later case we show examples that illustrate mapping of the matrix formulation into the concepts originating from the graph theory. Finally, attention is drawn to some novel aspects of the financial coherence which opens room for speculation if analogous effects can be detected in the atomic nuclei or in other strongly interacting Fermi systems.
\end{abstract}

\maketitle

\section{Complexity}

The concept of complexity refers to diversity of forms, to emergence of coherent patterns out of randomness and also to an extreme easiness of switching from one such pattern to another. Such properties typically accompany the systems that involve many degrees of freedom, many different space and time scales, and therefore such phenomena like chaos, noise, but also coherence, collectivity and criticality~\cite{Stanley}. It thus seems most appropriate to view the real complexity just at the interface of chaos and collectivity~\cite{Kauffman,Bak} as these two seemingly contradictory phenomena involve essentially the same elements. The recent concept of the scale free complex networks~\cite{Albert} offers one promising direction to formalize complexity though many related issues still remain open.

Studying complex systems, either empirically or theoretically, is unavoidably based on analyzing large multivariate ensembles of parameters. The most efficient starting formalism to grasp such ensembles and to quantify the whole variety of effects connected with complexity is in terms of matrices~\cite{Drozdz02}. Since complexity is embedded in chaos, or even noise, the random matrix theory (RMT)~\cite{Wigner,Mehta} provides then an appropriate reference. Its utility, at the first place, results from the fact that the degree of agreement quantifies the generic properties of a system - those connected with an overwhelming chaotic or noisy dynamics. For the complex systems this is expected to be a dominant component, but this component is not what constitutes an essence of complexity. From this perspective the deviations are even more relevant and more interesting as they may reflect constructive and perhaps deterministic components emerging from a noisy background.

The main related purpose of the present summary is to explore the financial characteristics of complexity in analogy to the ones that already are identified in strongly interacting quantum many body systems such as atomic nuclei. One such principal characteristics in the nuclear context is the energy separation of the collective state from the noisy background~\cite{Drozdz02,Stan98,Stan95}.

\section{Complex nuclei}

The structure of nuclei is a conventional example for collectivity embedded in chaos \cite{Stan98,Stan95}. The single particle structure in odd mass nuclei, the rotational spectra in deformed nuclei, the low-lying collective $2^+$ and $3^-$ states in spherical nuclei as well as the high-lying giant resonances are indications of a regular behavior in nuclei. The absolute number of such states in each nucleus, however, is small compared with the total number of levels. For that reason statistical methods have been developed and used to analyze the remaining huge number of individual levels \cite{Weidi}. The best example for collectivity embedded in chaos are giant resonances. These collective states exist in all medium and heavy mass nuclei. In Figure~1 the electric giant quadrupole resonance in several nuclei is shown. Collectivity means a cooperation, and thus the coupling between the different degrees of freedom in order to generate a coherent signal in response to an external perturbation. In shell model approaches collective states may be generated within the random phase approximation (RPA) in a 1p1h configuration space~\cite{Speth77}. For a quantitative description of the width of giant resonances, however, it is necessary to couple to the 2p2h space~\cite{Stan90} where, due to a sufficiently large density of states relative to the strength of the residual interaction, local level fluctuations characteristic for a Gaussian orthogonal ensemble (GOE) appear to already take place~\cite{Stan98,Stan94}. The collective 1p1h part of the giant resonances can thus be considered as embedded in the chaotic 2p2h space. The mechanism that creates coherent solutions can transparently be understood in a schematic model postulated by Brown and Bolsterli~\cite{Gerry}. Here one assumes that all 1p1h states are energetically nearly degenerate as indicated in the upper part of the right panel of Figure~1. As interaction one uses a separable multipole-multipole interaction which allows one to solve the RPA equation analytically. The result is shown in the lower part of the right panel of Figure~1. One state is strongly shifted in energy and includes all the transition strength $B(Q)$. All the other states loose nearly all their transition strength. In a more realistic description, if the 2p2h space is properly included, one is able to explain the experimental spectra shown in Figure~1~\cite{Stan90}.

\section{Coherence versus randomness in the matrix representation}

Expressed in the most general form, in essentially all the cases of practical interest, the $n \times n$ matrices ${\bf W}$ used to describe the complex system can be written as
\begin{equation}
{\bf W} = {\bf X} {\bf Y}^{\rm T},
\label{cab}
\end{equation}
where ${\bf X}$ and ${\bf Y}$ denote the rectangular $n \times m$ matrices. Such, for instance, are the correlation matrices whose standard form corresponds to ${\bf Y} = {\bf X}$. In this case one deals with $n$ observations or cases, each represented by an $m$ dimensional row vector ${\bf x}_i$ $({\bf y}_i)$, ($i=1,...,n$)~\cite{Muirhead}, and typically $m$ is larger than $n$. In the limit of purely random correlations the matrix ${\bf W}$ is then said to be a Wishart matrix~\cite{Wishart}. The resulting density ${\rho_{\bf W}(\lambda)}$ of eigenvalues is here known analytically~\cite{marcenko67,sengupta99}, with the limits $(\lambda_{min} \le \lambda \le \lambda_{max})$ prescribed by $\lambda^{max}_{min} = 1 + 1/Q \pm 2\sqrt{1/Q}$ and $Q = m/n \ge 1$:
\begin{equation}
\rho_{\rm W}(\lambda) = {Q \over 2 \pi \sigma^2} { \sqrt { (\lambda_{\rm max}-\lambda)(\lambda-\lambda_{\rm min}) } \over \lambda }.
\label{marcenko}
\end{equation}
This is the so-called Mar\v cenko-Pastur distribution. The variance $\sigma$ of the elements of ${\bf x}_i$ is in our case equal to unity.

\begin{figure}
\hspace{-1.0cm}
\includegraphics[height=.35\textheight]{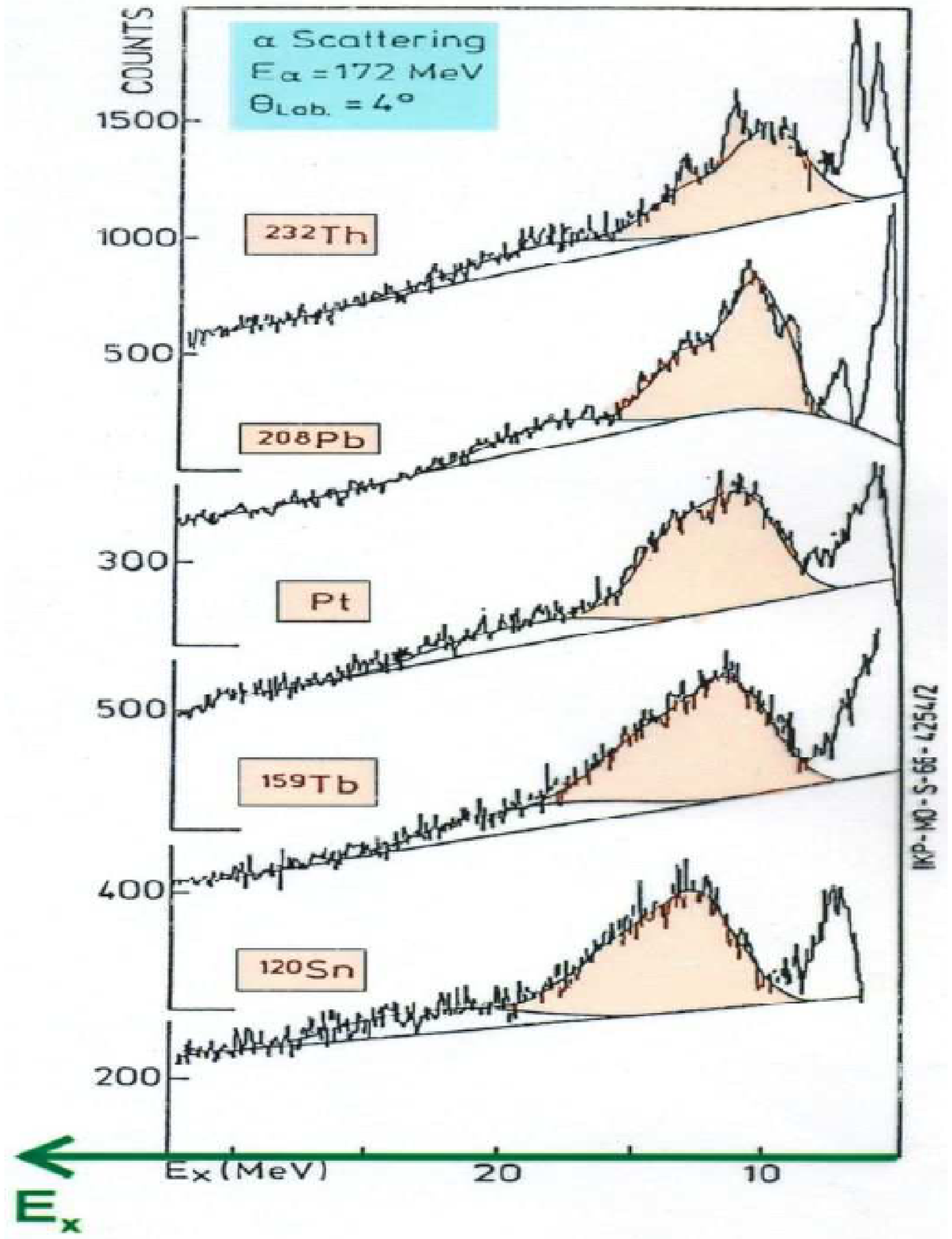}
\hspace{0.5cm}
\includegraphics[height=.35\textheight]{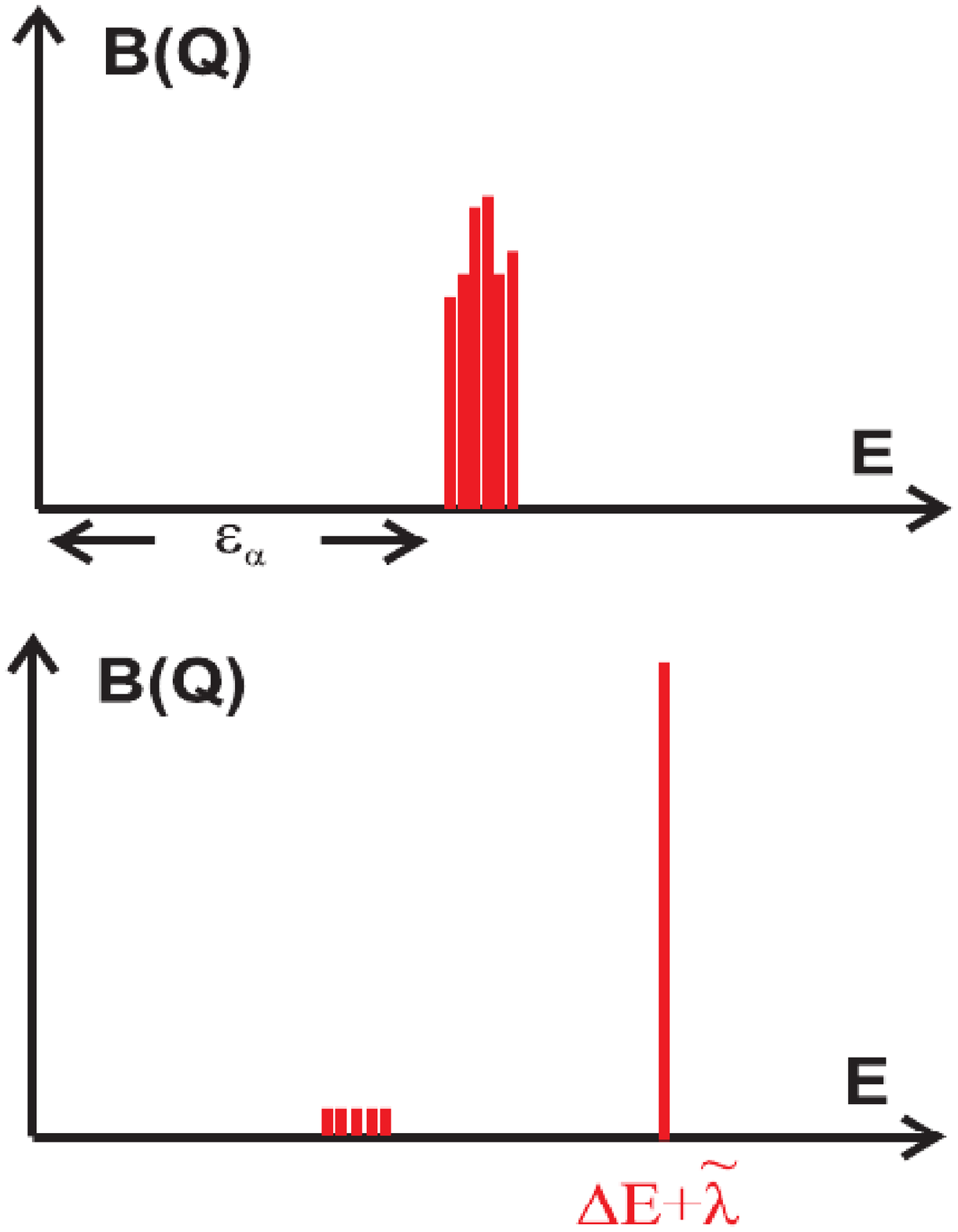}
\caption{Example from the nuclear physics. (Left) Giant quadrupole resonances in atomic nuclei. (Right) RPA solution in the schematic Brown-Bolsterli model; in the upper part the degenerate 1p1h states are shown, in the lower part the corresponding RPA solution for a repulsive ph-interaction.}
\end{figure}

As an appropriate reference for quantifying various characteristics of complexity one can use a few distinct types of random matrix ensembles, depending on a problem. In a typical situation, the bulk of such characteristics is consistent with RMT predictions. However, the genuine properties of a studied system are expected to manifest themselves in deviations from the universal pattern. These deviations are primarily connected with the reduction of dimensionality due to emergence of synchronous or coherent patterns. In the matrix formalism such patterns can be expressed with a significantly reduced rank of a leading component (or components) of ${\bf W}$. This can be written schematically as a decomposition:
\begin{equation}
{\bf W} = {\bf W}_{\rm RMT} + {\bf W}_{\rm coll} + \sum_i {\bf W}_i,
\label{ccc}
\end{equation}
in which the first component is the noisy (RMT) component, the second (usually dominating) component of rank 1 describes the coexisting collective global behavior of the system, while some possible additional components denoted by the sum are related to local coherent effects engaging only parts of the system.

In the absence of ${\bf W}_{\rm RMT}$, the other terms in (\ref{ccc}) generate $k$ nonzero eigenvalues, and all the remaining ones $(n-k)$ constitute the zero modes. The schematic Brown-Bolsterli model case discussed above corresponds to $k=1$ with no other terms in (\ref{ccc}). When - more realistically - ${\bf W}_{\rm RMT}$ enters as a noise correction, a trace of the above effect is expected to remain, i.e., $k$ large eigenvalues and the bulk composed of $n-k$ small eigenvalues whose distribution and fluctuations are (of course after disregarding the collective part) consistent with an appropriate random matrix ensemble. A likely mechanism that may lead to such eigenspectra is that $m$ in Eq.~(\ref{cab}) is actually significantly smaller than $n$, or that it is effectively small if the number of large ${\bf X}$ and ${\bf Y}$ entries is small.

Another mechanism that may lead to a structure analogous to (\ref{ccc}) is the presence - in addition to noise - of some systematic trend in the column vectors of ${\bf X}$ and ${\bf Y}$. Then~\cite{Drozdz02,Kwapien00}, to a first approximation, the ${\bf W}_{\rm coll}$ term is represented just by a matrix whose all entries are close in magnitude, and thus its rank is directly seen to be unity. This structure of a matrix can be easily detected if one observes an asymmetric shift of $P(w)$ relative to zero.

The matrix form as expressed by Eqs.~(\ref{cab}) and (\ref{ccc}) proves very general. The Hamiltonian matrices of strongly interacting quantum many body systems such as atomic nuclei can also be expressed in this form. This holds true~\cite{Stan01} on the level of bound states where the problem is described by the Hermitian matrices, as well as for excitations embedded in the continuum~\cite{Rotter}.

\section{Financial markets}

Financial markets have only relatively recently become area of non-episodic scientific interest for physicists. Despite some interesting earlier connections between physics and financial economics like the description of the price movements in terms of the random walks~\cite{fama65} or the approach to modeling of bond prices by means of the path integrals~\cite{ingber90}, physics and economics for decades were developing largely independently. This changed rapidly after financial institutions absorbed a number of physicists in 1990's and, at the same time, large amounts of computer-recorded financial data became easily available. Financial data are alluringly clean and this forms an additional factor attracting physicists and, among them, especially the statistical physicists.

From an empirical perspective, financial markets, like the stock markets or the currency markets, are systems with many degrees of freedom defined by the assets traded on a given market. Since price movements are inevitably dominated by trends that make data strongly nonstationary, it has become standard to analyze time series of price returns (defined as logarithmic price increments over fixed time intervals) instead of prices themselves. It does not solve the problem of nonstationarity completely, but at least allows one to successfully use the standard methods of time series analysis.

In accord with what has been presented in the previous sections, also a financial market can be described as a system of interacting or otherwise coupled elements-assets whose dependences can be expressed by a matrix. In the present case this role is played by a correlation matrix which, in accordance with Eq.~(\ref{cab}), is constructed from a data matrix ${\bf X}$ consisting of multivariate time series of returns of $n$ assets observed over equivalent time intervals. After diagonalization of the correlation matrix, one obtains its eigenspectrum which then can be compared with the RMT prediction for the Wishart ensemble.

\begin{figure}
\hspace{0.0cm}
\includegraphics[height=.3\textheight]{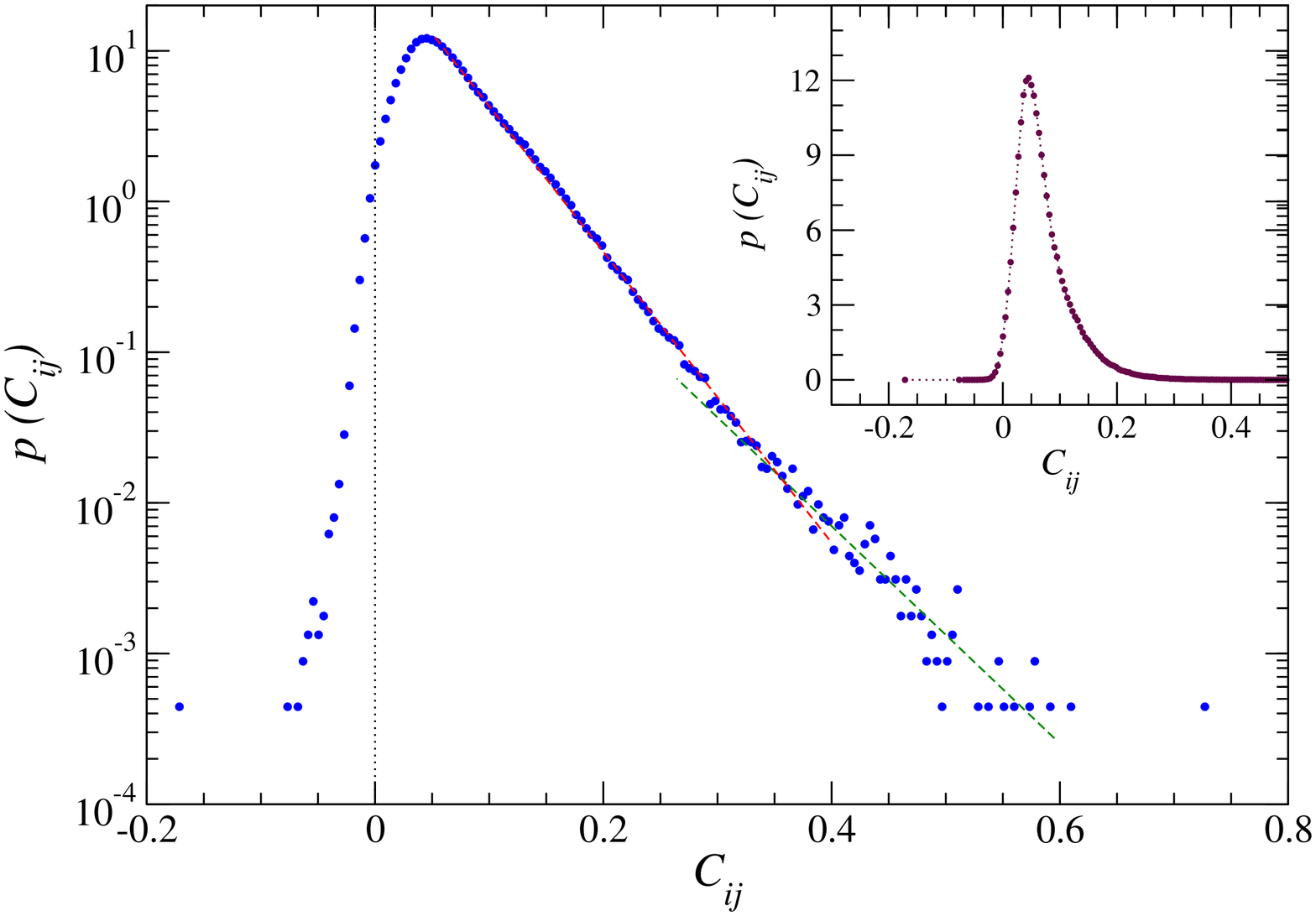}
\hspace{-0.5cm}
\includegraphics[height=.3\textheight]{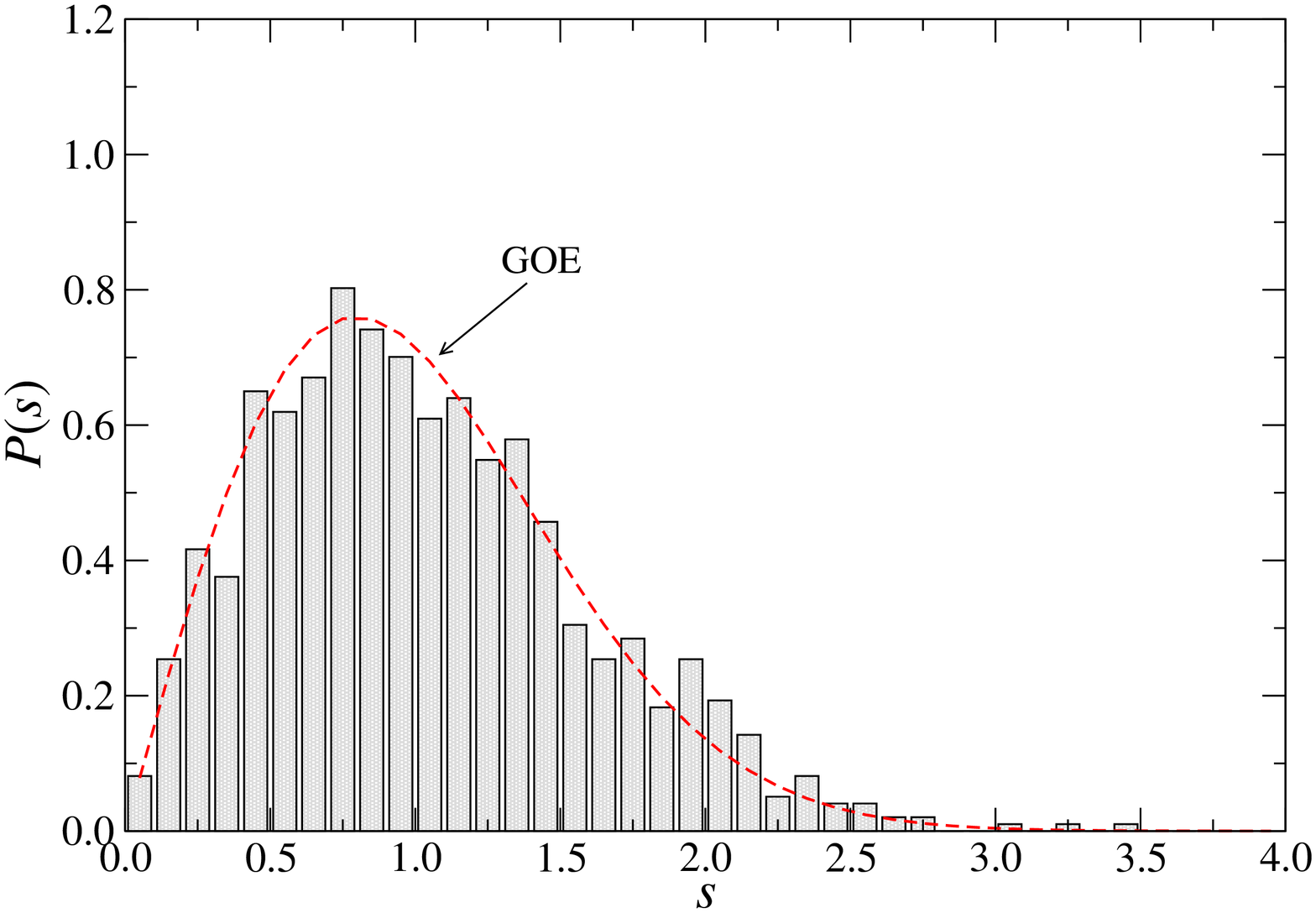}
\caption{Example from the financial physics. (Left) Distribution of correlation matrix elements for time series of 15 min. returns of 1000 American companies traded on NYSE and Nasdaq. Inset: the same but in the double linear scale. (Right) Eigenvalue spacings distribution for the same correlation matrix (histogram) compared with the Wigner distribution predicted for GOE matrices by RMT (dashed line).}
\end{figure}

As an illustration, we consider a set of time series of 15-min. returns for $n=1000$ highly-capitalized American companies traded on the New York Stock Exchange or on Nasdaq. Each time series is $m=13,546$ points long and comprises the interval 1998-1999. First of all, the left panel of Figure~2 shows the distribution of matrix elements which is the most straightforward measure of the matrix properties. The distribution is asymmetric and, exactly as discussed above, its center of mass is shifted towards positive values. As this Figure documents, the distribution is also far from Gaussianity.

What is now completely obvious, the eigenspectrum of our correlation matrix also deviates significantly from the RMT result~\cite{kwapien06}. This maybe is not so well-evident on the level of the eigenvalue spacings (right panel of Figure~2) whose distribution is roughly consistent with the Wigner distribution (the Wishart random ensembles do not differ in this respect from the Gaussian orthogonal ones). However, if one directly looks at the eigenvalue spectrum (left panel of Figure~3), there is very clear non-random structure of especially the largest several $\lambda$'s. The largest eigenvalue $\lambda_1=84.3$ is extremely strongly repelled from the rest of the spectrum. This eigenvalue can be identified with the ${\bf W}_{\rm coll}$ term in Eq.~(\ref{ccc}), which in the language of the financial markets is called ``the market factor'' and describes the collective movement of all assets (usually grasped by the behavior of a market composite index). A few smaller eigenvalues seen in Figure 3 have to be associated with the third term in Eq.~(\ref{ccc}); they typically correspond to market sectors (industries in the case of stock markets or geographical regions in the case of the currency market). It must be noted that the RMT bounds are quite narrow in the present case: $\lambda_{\rm min}=0.53$ and $\lambda_{\rm max}=1.62$.

\begin{figure}
\hspace{-0.5cm}
\includegraphics[height=.4\textheight]{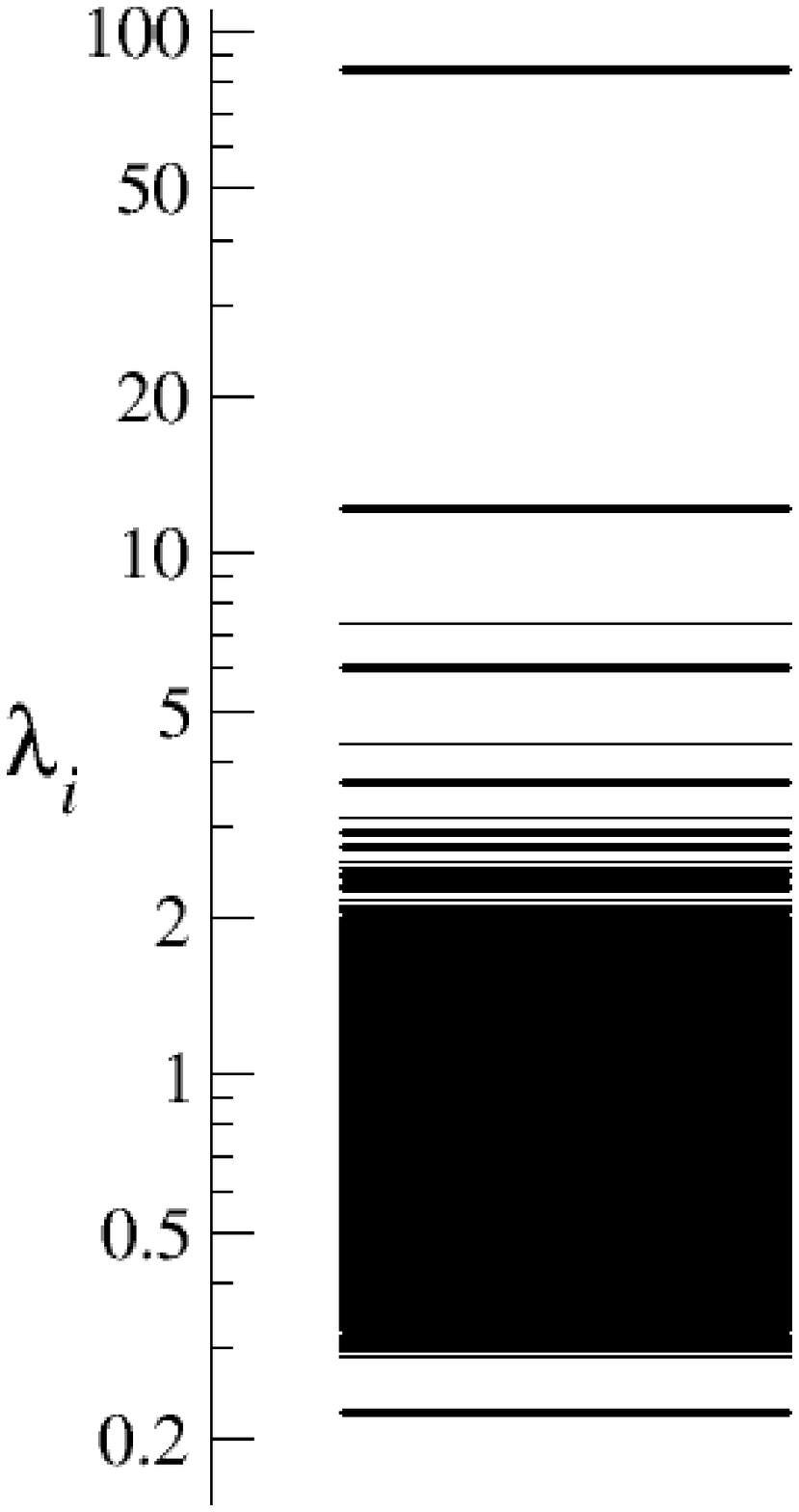}
\hspace{0.5cm}
\includegraphics[height=.4\textheight]{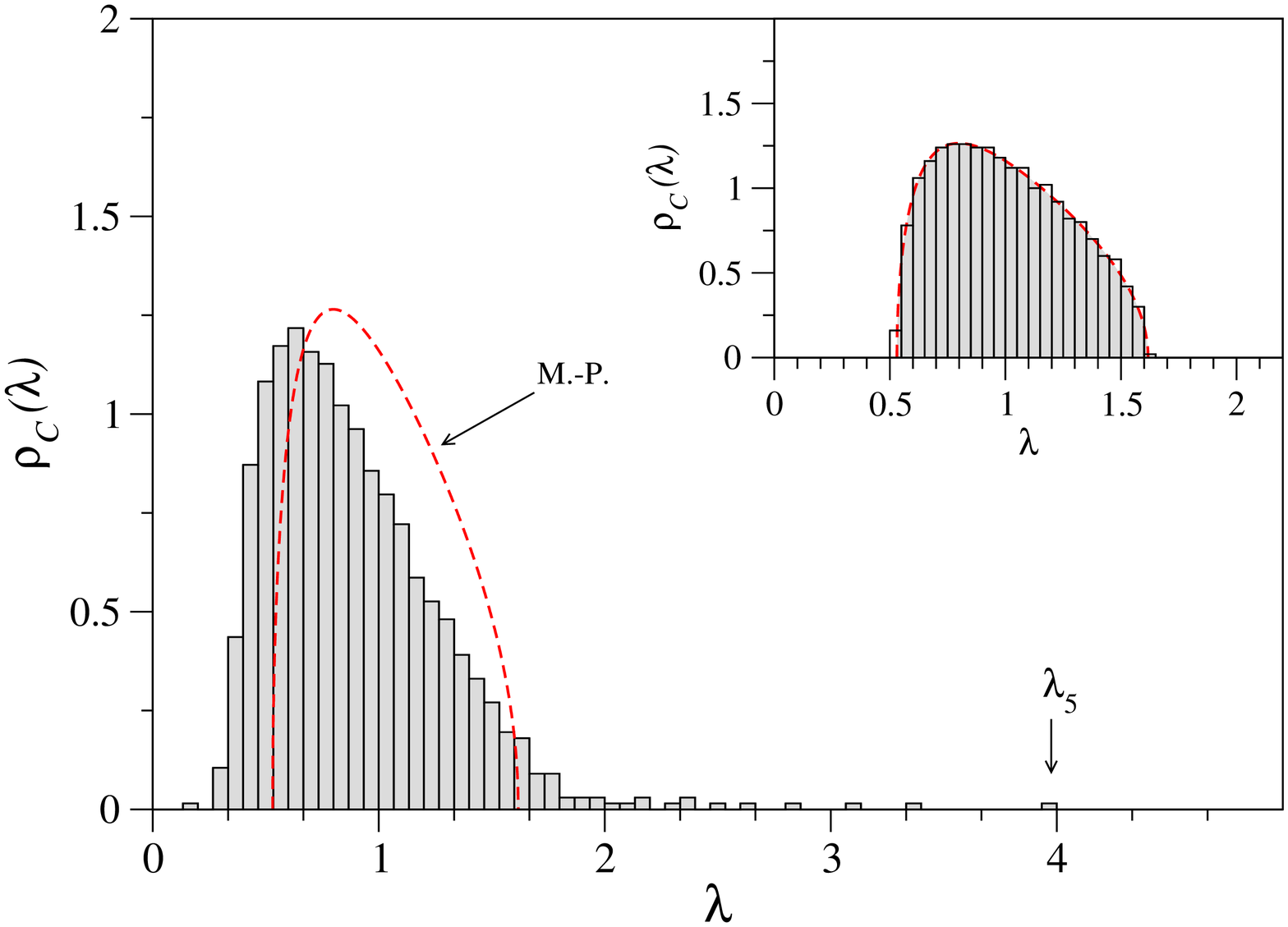}
\caption{(Left) Eigenvalues $\lambda_i$ of the correlation matrix calculated for time series of 15 min. returns of 1000 American companies traded on NYSE and Nasdaq. Note the logarithmic scale of the axis. (Right) Eigenvalue density for the same matrix (histogram) together with the corresponding Mar\v cenko-Pastur distribution predicted by RMT (dashed line). This empirical distribution does not show the four largest eigenvalues which, as it is seen in the left panel, are located beyond the range of the graph. Inset: Mar\v cenko-Pastur distribution in agreement with the eigenvalue distribution for a correlation matrix constructed from the randomized data (histogram).}
\end{figure}

A characteristics which allows one to quickly assess the existence of non-random components in the correlation matrix is the eigenvalue density. This is because of a concise result for the Wishart matrix ensemble expressed by (\ref{marcenko}). This theoretical distribution is shown in the right panel of Figure~3 together with the empirical eigenvalue distribution for the data under study. In fact, the discrepancies between the two are significant both at the ends and in the center. The eigenvalue density for the randomized time series (surrogates) is seen in the inset. This proves that the non-random properties of the actual eigenvalue density come from the inter-stock correlations destroyed by the shuffling of data.

\begin{figure}
\hspace{-0.5cm}
\includegraphics[height=.35\textheight]{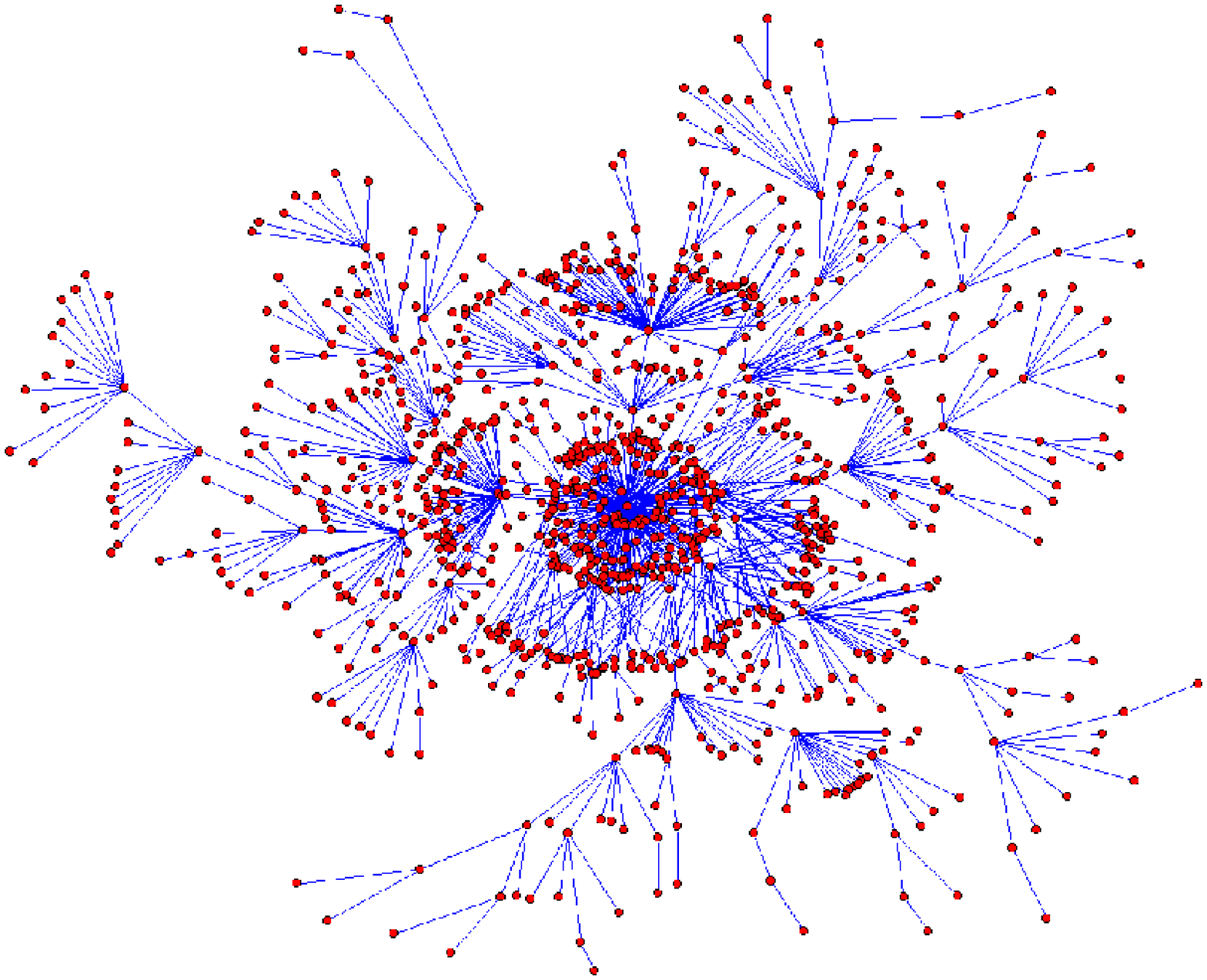}
\hspace{0.0cm}
\includegraphics[height=.35\textheight]{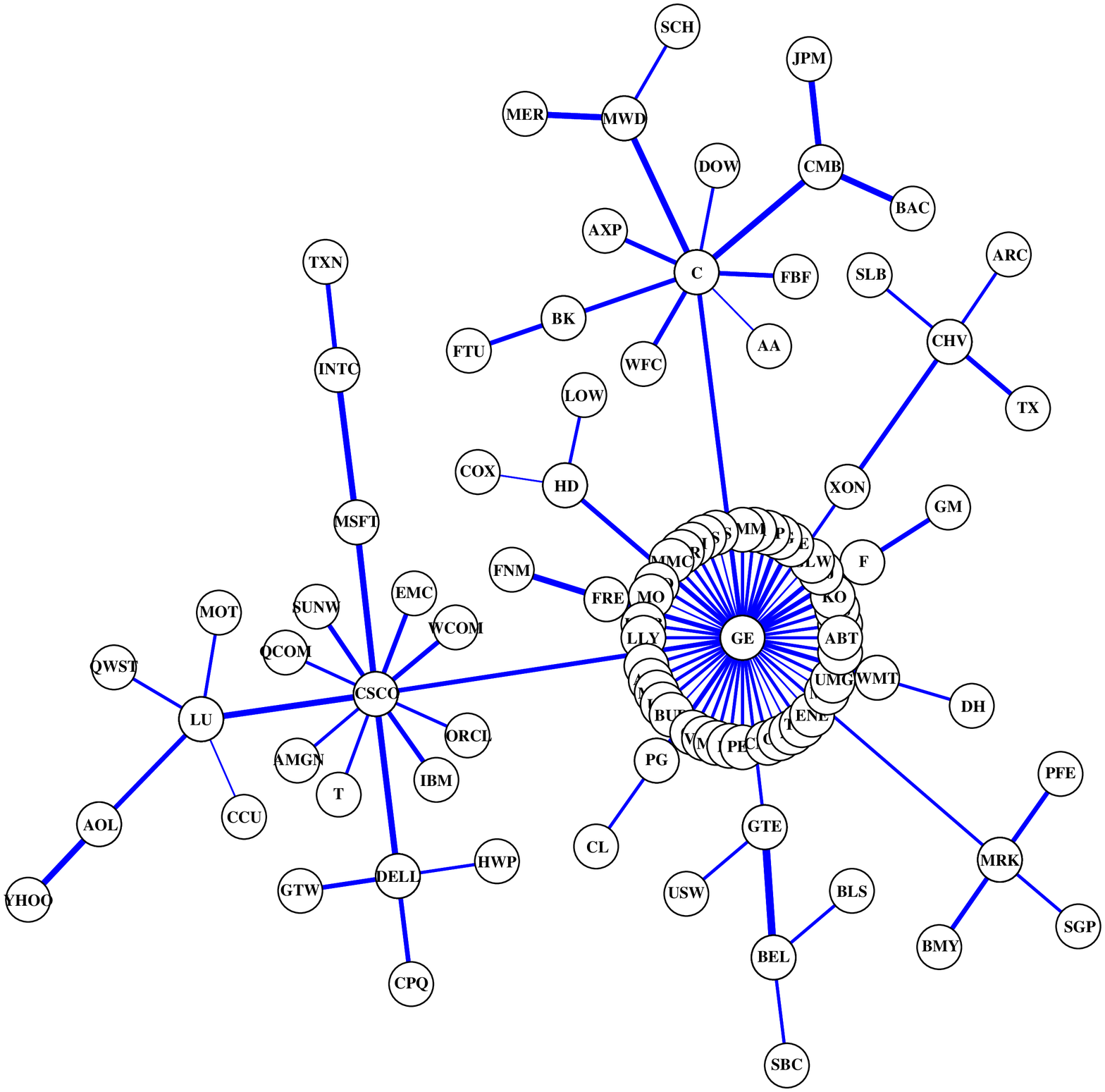}
\caption{(Left) Minimal spanning tree for a set of 1000 large American companies traded in NYSE and Nasdaq. (Right) MST for a subset of 100 American companies of the largest capitalization; edge widths are proportional to the corresponding metric distances~(\ref{dist}).}
\end{figure}

The results presented so far describe the statistical properties of dependences between the assets but do not allow us to visualize the exact microscopic structure of these dependences. This may be considered a drawback especially in the case of financial markets where, for practical reasons, the details are important. This issue can, however, be overcome with the network-oriented approach, in which the relevant relations among the system's elements can be expressed by a structure of nodes and edges. In order to illustrate the point, we create a network representation of our set of 1000 American stocks. Each asset is defined as a node and each pair of assets is connected by a link with a weight equal to the corresponding correlation matrix entry. However, the so-defined network consists of 1000 nodes and 495,000 edges which makes it virtually impossible to be presented graphically. Therefore we have to confine our task to the construction of a sub-network of ``minimum weight'', i.e. the sub-network in which connections between nodes of each pair are the strongest possible. This is the so-called minimum spanning tree (MST).

The MST is calculated based on the correlation matrix after transforming its entries $w_{ij}$ into metric distances~\cite{mantegna99} according to the following relation:
\begin{equation}
d_{ij} = \sqrt{ 2 (1 - w_{ij}) }.
\label{dist}
\end{equation}
The distances $d_{ij}$ are sorted from the smallest to the largest forming a list and by going from the top of this list one connects pairs of nodes which have not yet been connected to each other by any continuous path. In this way, after passing through the whole list, one finally obtains a dendrogram with $n$ nodes and $n-1$ edges, which is relatively easy to draw and analyze. MST for our set of assets is displayed in Figure~4 (left panel). It is striking that the resulting network consists of a central group of few stocks to which all other stocks, in a form of layers, are connected. This can be interpreted as an existence of companies which are treated by investors as the whole American economy in a nutshell. Other companies are thus viewed as peripheral and less representative.

In order to better see the core structure of the market and identify the ``hubs'', the same procedure was applied to the subset of 100 companies with the largest market capitalization (roughly $> 3 \cdot 10^{10}$ USD). The corresponding MST is shown in the right panel of Figure~4. Now it is clear that the most important node is General Electric (GE) and it is accompanied by less central, but still important, nodes of Citigroup (C) and Cisco Systems (CSCO). GE is a huge conglomerate with diversified areas of activity and this fully justifies its dominant role in the network. On the other hand, C is a part of a large industry of financial institutions and CSCO belongs to the high-tech market sector which is also significant. Other companies which share the entire or a part of activity areas of the dominant companies (or are otherwise related) are naturally connected to them by direct links, forming the inner layers. In the same manner, the rest of nodes form the outer layers since their profiles significantly differ from the profiles of the three hubs.

This example of MSTs shows that if one needs to grip the key features of the microscopic structure of a system quickly and in a concise manner, the network formalism proves very useful~\cite{gorski08,kwapien09}. But also more general and global characteristics of dependences among the system's constituents can be described using approaches based on the networks theory. This refers primarily to the case of complex systems in which complexity manifests itself by such features like scale invariance or the small world structure~\cite{albert99,watts98} and the matrix representation offers an appropriate starting point.  

\section{Summary}

As shown above, several concepts originally developed in the area of nuclear physics and the related experience of handling complexity prove useful in attacking even more complex systems such as the financial markets. Some global characteristics like the mechanism of the coexistence of chaos and collectivity appear to be similar. As far as the financial collectivity is concerned there are some novel elements however. One of the related distinct patterns, likely originating from the hierarchical organization illustrated in the previous Section, are imprints of the discrete scale invariance which manifest themselves via the log-periodic oscillations~\cite{Sornette03} cascading self-similarly through various time scales~\cite{Drozdz99} with amazingly universal quantitative characteristics~\cite{Drozdz03,Bartolozzi}. It is not excluded that at a sufficiently high energy resolution the analogous effects can be detected also in the atomic nuclei. Finally, the insight offered by the network representation, as above illustrated for the stock market, may encourage attempts to graphically represent the nuclear Hamiltonian as well, in the mean field basis for instance.


\begin{thebibliography}{99}

\bibitem{Stanley} H.E.~Stanley, {\it Rev. Mod. Phys.} {\bf 71} S358 (1999).
\bibitem{Kauffman} S.A.~Kauffman, {\it The Origins of Order--Self-Organization and Selection in Evolution}, Oxford University Press, Oxford, 1993.
\bibitem{Bak} P.~Bak, {\it How Nature Works--the Science of Self-Organized Criticality}, Springer-Verlag, New York, 1996.
\bibitem{Albert} R.~Albert, A.-L.~Barabasi, {\it Rev. Mod. Phys.} {\bf 74} 47 (2002).
\bibitem{Drozdz02} S.~Dro\.zd\.z, J.~Kwapie\'n, J.~Speth, M.~W\'ojcik, {\it Physica A} {\bf 314} 355 (2002).
\bibitem{Wigner} E.P.~Wigner, {\it Ann. Math.} {\bf 53} 36 (1951).
\bibitem{Mehta} M.L.~Mehta, {\it Random Matrices}, Academic Press, Boston, 1991.
\bibitem{Stan98} S.~Dro\.zd\.z, S.~Nishizaki, J.~ Speth, M.~W\'ojcik, {\it Phys.Rev E} {\bf 57} 4016 (1998).
\bibitem{Stan95} S.~Dro\.zd\.z, S. Nishizaki, J. Wambach, J. Speth. {\it Phys. Rev. Lett.} {\bf 74} 1075 (1995).
\bibitem{Weidi} T.~Guhr, A.~M\"uller-Groeling, H.A.~Weidenm\"uller, {\it Phys. Rep.} {\bf 299} 189 (1998).
\bibitem{Speth77} J.~Speth, E.~Werner,  W.~Wild, {\it Phys. Rep.} {\bf 33} 127 (1977)
\bibitem{Stan90} S.~Dro\.zd\.z, S.~Nishizaki, J.~Speth, J.~Wambach, {\it Phys. Rep.} {\bf 197} 1 (1990).
\bibitem{Stan94} S.~Dro\.zd\.z, S.~Nishizaki, J.~Speth, J.~Wambach, {\it Phys. Rev. C} {\bf 49} 867 (1994).
\bibitem{Gerry} G. E.~Brown, M.~Bolsterli, {\it Phys. Rev. Lett.} {\bf 3} 479 (1959).
\bibitem{Muirhead} R.J.~Muirhead, {\it Aspects of Multivariate Statistical Theory}, John Wiley $\&$ Sons, New York, 1982.
\bibitem{Wishart} J.~Wishart, {\it Biometrica} {\bf 20} 32 (1928).
\bibitem{marcenko67} V.A.~Mar\v cenko, L.A.~Pastur, {\it Math.~USSR~Sb.} {\bf 1}, 457-483 (1967).
\bibitem{sengupta99} A.M.~Sengupta, P.P.~Mitra, {\it Phys. Rev. E} {\bf 60} 3389 (1999).
\bibitem{Kwapien00} J.~Kwapie\'n, S.~Dro\.zd\.z, A.A.~Ioannides, {\it Phys. Rev. E} {\bf 62} 5557 (2000).
\bibitem{Stan01} S.~Dro\.zd\.z, M.~W\'ojcik, {\it Physica A} {\bf 301} 291 (2001).
\bibitem{Rotter} I.~Rotter, {\it Rep. Prog. Phys.} {\bf 54} 635 (1991).
\bibitem{fama65} E.F..~Fama, {\it Fin.~Anal.~J.} {\bf 21} 55--59 (1965).
\bibitem{ingber90} L.~Ingber, {\it Phys.~Rev.~A} {\bf 42} 7057--7064 (1990).
\bibitem{kwapien06} J.~Kwapie\'n, S.~Dro\.zd\.z, P.~O\'swi\c ecimka, {\it Physica A} {\bf 359} 589--606 (2006).
\bibitem{mantegna99} R.N.~Mantegna, {\it Eur.~Phys.~J.~B} {\bf 11}, 193-197 (1999).
\bibitem{gorski08} A.Z.~G\'orski, S.~Dro\.zd\.z, J.~Kwapie\'n, {\it Eur.~Phys.~J.~B} {\bf 66}, 91-96 (2008).
\bibitem{kwapien09} J.~Kwapie\'n, S.~Gworek, S.~Dro\.zd\.z, {\it Acta Phys.~Pol.~B} {\bf 40}, 175--194 (2009).
\bibitem{albert99} R.~Albert, H.~Jeong, A.-L.~Barab\'asi, {\it Nature} {\bf 401}, 130--131 (1999).
\bibitem{watts98} D.J.~Watts, S.H.~Strogatz, {\it Nature} {\bf 393}, 440--442 (1998).
\bibitem{Sornette03} D. Sornette, \textit{Why Stock Markets Crash: Critical Events in Complex Financial Systems}, (Princeton University Press, Princeton, 2003).
\bibitem{Drozdz99} S.~Dro\.zd\.z, F.~Ruf, J.~Speth, M.~W\'ojcik, {\it Eur. Phys. J. B} {\bf 10}, 589 (1999).
\bibitem{Drozdz03} S.~Dro\.zd\.z, F.~Gr\"ummer, F.~Ruf, J.~Speth, {\it Physica A} {\bf 324}, 174 (2003).
\bibitem{Bartolozzi} M.~Bartolozzi, S.~Dro\.zd\.z, D.~B.~Leinweber, J.~Speth, A.~W.~Thomas, {\it Int. J. Mod. Phys. C} {\bf 16}, 1347 (2005).

\end{thebibliography}
\end{document}